\documentclass[aps,twocolumn,superscriptaddress,showpacs]{revtex4}

\usepackage{graphicx}
\usepackage{amsfonts}
\usepackage{amssymb,amsmath}

\begin{document}
\title{Negativity and quantum discord in Davies environments}
\author{J. Dajka}
\address{Institute of Physics, University of Silesia, 40-007 Katowice, Poland }
\author{M. Mierzejewski}
\address{Institute of Physics, University of Silesia, 40-007 Katowice, Poland }
\author{J. {\L}uczka}
\address{Institute of Physics, University of Silesia, 40-007 Katowice, Poland }
\author{R. Blattmann}
\address{Institute of Physics, Augsburg University, Germany}
\author{P. H{\"a}nggi}
\address{Institute of Physics, Augsburg University, Germany}
\begin{abstract}
We investigate the  time evolution of   negativity  and  quantum discord for a pair of non-interacting qubits with one being weakly  coupled to a decohering Davies--type Markovian environment.  At initial time of preparation, the  qubits are prepared in one of the maximally entangled pure Bell states. In the  limiting case of pure decoherence (i.e. pure dephasing), both, the  quantum discord and negativity decay to zero in the long time limit. In presence of a manifest dissipative dynamics,  the entanglement negativity undergoes a sudden death at finite time while the quantum discord relaxes continuously to zero with increasing  time.
We find that in dephasing environments the decay of the negativity is  more propitious  with increasing time; in contrast, the evolving decay of the quantum discord proceeds weaker for dissipative environments. Particularly,
the slowest  decay of the quantum discord emerges when the  energy relaxation time  matches the  dephasing  time.
\end{abstract}

\maketitle
%
\section{Introduction}

Establishing correlations is a {\it sine qua non} condition in  effective  communication. If two parties are {\it quantum correlated} one can attempt to conduct a quantum communication between them. Quantum communication protocols \cite{ent} make use of certain properties of states of quantum composite systems.
The best known and  most popular resource for quantum communication is quantum entanglement \cite{hor}, a widely studied property of composite systems. Quantum entanglement \cite{hor}, interesting {\it per se}, has been recognized as a powerful tool for quantum information processing \cite{ent}. For example, let us mention  the use of quantum communication with dense coding and teleportation as most celebrated examples \cite{hor,ent}.



If one presupposes that the history of quantum entanglement started out with the work by Schr\"{o}dinger \cite{sch}, it took almost a century to discover that there are quantum correlations which are essentially non-classical and these can exist even in the absence of entanglement \cite{zur,discord}. When compared to entanglement, quantum correlations are, to some extent, 'more mysterious', since their   mathematical setting  has not been uniquely established. The well known mathematical language for entanglement incorporating tensor products of state spaces, Schmidt decomposition, etc.,  not only allows one to pose many of fundamental problems in purely algebraic context but also motivates the search for a variety of entanglement quantifiers.
The physical context of entanglement encoded e.g. in Bell--type inequalities \cite{hor} can be  (sometimes incorrectly)  considered as a consequence of mathematical theorems.  Recent studies of quantum correlations  have shown that a broad class of composite systems carry correlations which can be described in the context of information extracted from composite parties via a suitably formalized measurement procedure, see in Ref. \onlinecite{discord} for recent review.

Among several quantifiers of quantum correlations  the {\it quantum discord} attained recent popularity \onlinecite{discord}. Despite its  fundamental meaning, exemplarily studied  via  approximating a reduced quantum dynamics within  trace preserving completely positive  maps \cite{rosario},  quantum correlations quantified by quantum discord can open new avenues for quantum computations \cite{comp} and quantum communication schemes \cite{resource,merg,broad}.

A salient obstruction in the  implementation of both quantum  computation and communication protocols is quantum  decoherence \cite{ent}, induced  by the influence of the environment on the relevant 'quantum hardware'.  Studies of entanglement dynamics in the presence of noise enjoy a long history \cite{flor,eberly}; in contrast, similar studies on quantum discord dynamics have been carried out only recently  \cite{dys,disdyn,ali1,luo}.

According to Ref. \onlinecite{ali}, the
quantum discord and the entanglement present fundamentally different resources. Here, we investigate if this difference manifests itself in their robustness to  environment--induced decoherence. In other words, we study whether it is possible to assign an environment to two general classes, one of them more suitable for quantum information processing using entanglement--based protocols, and the second preferring quantum correlations as being quantified by quantum discord. With this study we obtain  an affirmative answer to this question.

In doing so we consider a simple set-up: We investigate the time evolution of quantum correlations between a pair of qubits with  only one of them being coupled to a decohering environment. We emphasize here that the {\it exact} reduced quantum dynamics for the two qubits is typically neither completely positive \cite {Pechukas, Sudarshan}, nor is it generally even linear \cite{Romero}, not to speak of undergoing a memoryless quantum Markovian dynamics \cite{NM}.  Here, however, we restrict ourselves to a quantum Markovian dynamics, having in mind quantum optics with weak coupling at not too low temperatures. Then the time evolution can satisfactorily be approximated with   a  trace preserving completely positive Markovian map of Davies-type \cite{alicki}.

The structure of the work is as follows: In Sec.  2, we
describe our model set-up and its dynamics in terms of a Davies map.  Next, in Sec.  3, we consider the time evolution of the correlation quantifier negativity while with Sec. 4, we analyze the quantum discord.
With Sec. 5 we present our  summary and  conclusions.

\section{Open quantum dynamics: Quantum Markovian Davies map} \label{secII}

With this work, somewhat stimulated by prior work  done in  Ref. \onlinecite{disdyn}, we consider a pair of non-interacting qubits $A$ and $B$  that   initially are prepared in one of the maximally entangled pure Bell states; i.e.,
\begin{eqnarray}\label{bells}
\rho_i &=& \left(\sigma_0\otimes\sigma_i\right)\rho_0\left(\sigma_0\otimes\sigma_i\right), i=0,1,2,3\\
\sigma_0&=&{\bf 1},\,\, \sigma_{1,2,3}=\sigma_{x,y,z}, \nonumber
\end{eqnarray}
where
\begin{eqnarray}\label{bell0}
\rho_0&=&\frac{1}{2}\left( |01\rangle+|10\rangle\right)\left( \langle 01|+\langle 10|\right).
\end{eqnarray}
 The non--decohering  part of the  evolution of the qubits $A$ and  $B$  is determined  by the two  two-level Hamiltonians; i.e.,
\begin{eqnarray}\label{ham}
H_A=\frac{\omega_A}{2} \left(\begin{array}{cc}1 &0\\0&-1  \end{array}\right),
\quad
H_B=\frac{\omega_B}{2}\left(\begin{array}{cc}1 &0\\0&-1  \end{array}\right).
\end{eqnarray}
%
For the following, only  qubit $B$  is coupled at times $t>0$ to an environment $E$ at temperature $T$.  Then, in the presence of this environment coupling,  the  dynamics of the two qubits, described by the reduced density matrix $\rho_{AB}(t)$,   no longer proceeds unitary. Because the two qubits $A$ and $B$ are initially correlated, both the quantum entanglement and the quantum discord will evolve in course of evolving time $ t > 0$,  as these become influenced by a finite  system $B$-environment coupling.

\begin{figure}[]
\includegraphics[width=0.5\textwidth,angle=0]{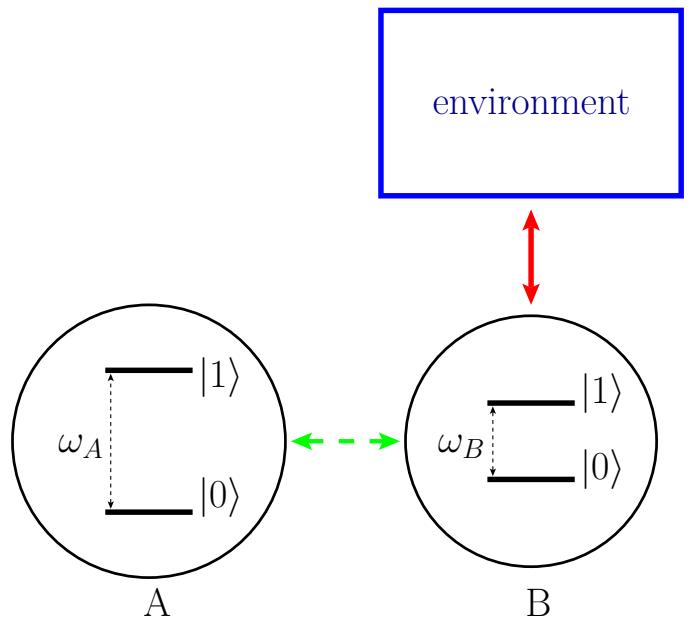}
\caption{(Color online) The sketch of the system set-up studied with this work. The qubits $A$ and $B$ with level separation $\omega_A$ and $\omega_B$, respectively, do physically not interact;  these are, however,  initially prepared in one of the maximally entangled  Bell states, as indicated by the broken (green) line. Here, only the qubit $B$ is weakly coupled to an environment. The  dynamics of the qubits is given by Eq. (\ref{evol}) in the text.}
\label{fig1}
\end{figure}

In presence of a very weak system--reservoir interaction and not extremely low temperatures the deduced dynamics can satisfactorily be described \cite{NM} by a Markovian dynamics of the Davies type \cite{alicki}. The main advantage of this approach is that it  recovers the well established steady state properties at assumed weak coupling, namely  stationarity and asymptotic stability in terms of the  Gibbs state, reading
\begin{eqnarray}\label{gibbs}
\rho_B(t)\rightarrow\left(\begin{array}{cc}p & 0\\0& 1-p \end{array}\right) \,\, \mbox{for} \, \, t\rightarrow \infty \;.
\end{eqnarray}
%
Here, the thermal weight is $p=\exp(-\beta \omega_B)/{\cal Z}$,
${\cal Z}=\exp(-\beta \omega_B) + \exp(\beta \omega_B)$ and $\beta = 1/kT$ denotes the inverse temperature.
Using  Davies theory one explicitly can construct the generator of a completely positive semigroup describing the reduced dynamics (with respect to the environment) of the open quantum system in terms of microscopic parameters in the Caldeira--Leggett-type Hamiltonian of the full system.
Such modeling has recently been applied to studies of entanglement dynamics \cite{lendi}, properties of a dissipative geometric phases of qubits \cite{dav_faza} and for some thermodynamic properties of nano-scale systems \cite{dav_heat}, to name but a few.

Here, instead of using the most general scenario of Davies semigroups, we restrict ourselves to a completely positive Davies map \cite{dav}. The  qubit--qubit reduced dynamics is given by the following Davies map
$\Phi_t^B$  \cite{dav}, acting on the Hilbert space $B$ and completed with the   tensorized unitary time evolution  for the uncoupled qubit $A$, the latter acting solely on the Hilbert space of qubit $A$. Put differently, we have for the reduced dynamics a (super-operator)-time evolution, reading
\begin{eqnarray}\label{evol}
\rho_{AB}(t)= \left( \mathbb{U}^A_t \otimes\Phi_t^B \right) \rho_i, \quad i=0,1,2,3.
\end{eqnarray}
This is because the $A$-system Hamiltonian  $H_A$ commutes with all other remaining Hamiltonians of the system $B$ and environment $E$. Stated explicitely,  for any    linear combination of matrices the form $|i_A\rangle\langle j_A|\otimes |i'_B\rangle\langle j'_B|$, we have   $\left( \mathbb{U}_t^A \otimes\Phi_t^B \right) |i_A\rangle\langle j_A|\otimes |i'_B\rangle\langle j'_B|= \mathbb{U}_t^A (|i_A\rangle\langle j_A|) \otimes \Phi_t^B (|i'_B\rangle\langle j'_B|)$.
The part of the super-operator  $\mathbb{U}_t$ is the unitary evolution operator generated by the Hamiltonian $H_A$; i.e.,  $\mathbb{U}_t (|i_A\rangle\langle j_A|) = \mbox{e}^{-iH_A t} |i_A\rangle\langle j_A| \mbox{e}^{iH_A t}$.
Let us describe how the non-unitary Davies map $\Phi_t^B$ acts on the one--qubit density matrix $\rho_B$. Following \cite{dav} let us construct a super--operator $\mathbf{\Phi}_t^B$ corresponding to  $\Phi_t^B$ acting on vectorized matrices
\begin{eqnarray}
\rho_B = \left(\begin{array}{cc}\rho_{00} & \rho_{01}\\ \rho_{10}& \rho_{11} \end{array}\right)
&\rightarrow& ||\rho_B\rangle\rangle=\left(\begin{array}{c} \rho_{00}\\\rho_{01}\\\rho_{10}\\\rho_{11}  \end{array}\right), \\
\Phi_t^B \; \rho_B &\rightarrow & \mathbf{\Phi_t^B} \;  ||\rho_B\rangle\rangle.
\end{eqnarray}
An explicit form of the  Davies super--operator then
reads \cite{dav}:   
\begin{eqnarray}\label{Phi}
\mathbf{\Phi_t^B}=\left(\begin{array}{cccc} 1-u(t) & 0& 0& u(t)p/(1-p) \\
                               0 &v(t) &0  &0 \\
                               0& 0& v^*(t) &0 \\
                               a &0 &0 &1-u(t)p/(1-p) \end{array} \right)
\end{eqnarray}
with the corresponding  relaxation functions reading
\begin{eqnarray}\label{pars}
u(t) &=&(1-p)[1-\exp(-F t)], \\
v(t) &=& \exp(-i\omega_B t-G t).
\end{eqnarray}
The parameters $F=1/\tau_1$ and $G= 1/\tau_2$ are related to the energy relaxation time $\tau_1$ and  the dephasing time $\tau_2$, respectively. Given the fact that these relaxation times are subjected to obey  the physical inequalities \cite{T12}
\begin{eqnarray}\label{warun}
G &\ge& F/2 \ge 0,  \quad \mbox{i.e.} \quad  2 \tau_1 \ge \tau_2,
\end{eqnarray}
it then follows that this Davies map indeed is a trace-preserving, completely positive map.
\begin{figure}[]
\includegraphics[width=0.45\textwidth,angle=0]{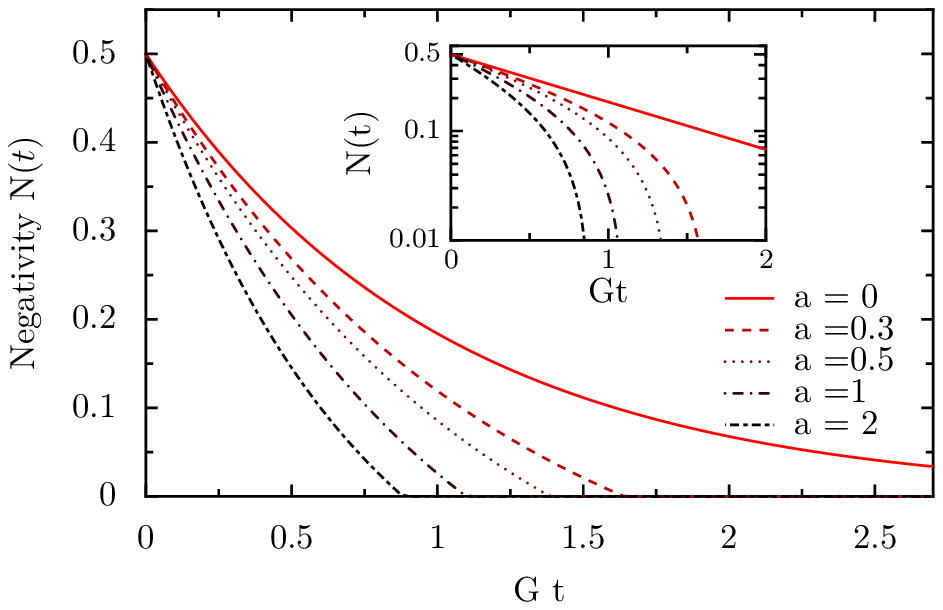}
\includegraphics[width=0.45\textwidth,angle=0]{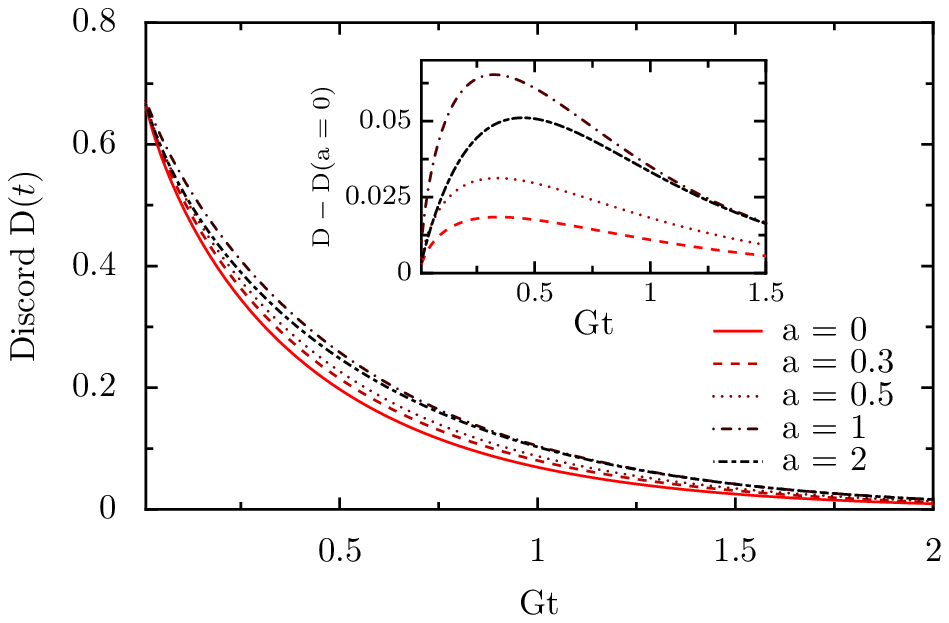}
\caption{(Color online) Upper panel: The entanglement quantifier for negativity  $N(t)$ [see Eq. (\ref{negat})] at temperature $T=\infty$, i.e. $p=1/2$, as a function of the dimensionless time $Gt$ for several  values of the ratio  $a=F/G=\tau_2/\tau_1$, where $\tau_2$ and $\tau_1$ are the dephasing and the energy relation times, respectively.
The inset depicts the non-exponential decay of negativity $N(t)$ for a dissipative environment, implying  that   $F>0$ (yielding  $a> 0$).  The negativity  $N(t)$ {\it vs.} the dimensionless time $Gt$ is depicted on the semi-logarithimic plot with the vertical axis  scaled logarithmically.  Lower panel: Time evolution of the quantum discord $D(t)$,  [see Eqs. (\ref{dys_0}) and (\ref{dys_pi})].  The inset shows the difference between discord in the dissipative case(i.e. $F>0$) and for pure
 dephasing (i.e. $F=0$) for several values of the ratio $a$. The slowest decay of quantum discord $D(t)$ is observed when $\tau_1=\tau_2$. }
\label{fig2}
\end{figure}

Let us consider the limiting case of an infinite energy relaxation rate, yielding $F=0$. This corresponds to pure dephasing without relaxation in energy taking place (no dissipation).  Such dephasing scenarios, including also non-Markovian  dephasing models, have been applied to studies of entanglement \cite{my} and quantum discord \cite{fini} dynamics. The opposite case, i.e. $G=0$ and $F\ne 0$ can physically not be realized: If there is dissipation of energy then necessarily finite dephasing accompanies this relaxation process.
For $p=1/2$ in Eq.(\ref{gibbs}),  i.e. in the case of an infinite large temperature, the  single qubit $B$  state becomes  maximally mixed:  $\mbox{Tr}_A(\rho_{AB}(t)) = \frac {1}{2} {\mathbb{I_B}}$. Likewise, we have for qubit A that $=\mbox{Tr}_B(\rho_{AB}(t))= \frac{1}{2} {\mathbb{I_A}}$. Hence with  $p=1/2$ the quantum discord, being evaluated below, is symmetric with respect to $A$ and $B$ labelling \cite{discord}.

Because the explicit results below are {\it robust} with respect to any chosen value for the Boltzmann weight  $p$, we shall depict in our Figures the  case with  $p=1/2$ only, yielding a symmetric quantum discord.

Given that at initial time  the reduced  two-qubit density matrix $\rho_{AB}(0) =\rho_0$,  where $\rho_0$ is the Bell state given by Eq. (\ref{bell0}), the density matrix $\rho_{AB}(t)$  evolving under the  Davies map (\ref{Phi})  assumes  the general time-dependent main result:
%
%
\begin{eqnarray}\label{rho}
& &   \rho_{AB}(t) = \nonumber \\
 \ \ \ \nonumber \\
&=& \frac{1}{4} \left( \begin {array}{cccc} 1-{{\rm e}^{-Ft}}&0&0&0
\\\noalign{\medskip}0&1+{{\rm e}^{-Ft}}& 2{{\rm e}^{i \omega t}}{
{\rm e}^{-Gt}}&0\\\noalign{\medskip}0&2{{\rm e}^{-i \omega t}}{{\rm e}^{-
Gt}}&1+{{\rm e}^{-Ft}}&0\\\noalign{\medskip}0&0&0&1-{
{\rm e}^{-Ft}}\end {array} \right), \nonumber \\
\ \ \
\end{eqnarray}
%
%
where $\omega = \omega_A - \omega_B$ and $\rho_{AB}(t=0) = \rho_0$.  This density matrix assumes the form of a so termed  $X$--state \cite{Yu,ali}. Note that the $X$--structure   of the reduced density matrix (\ref{rho}) is preserved during  time-evolution. This feature originates both from the symmetry of initial
preparation  (\ref{bells}), here assumed to be given by Eq.(\ref{bell0}), and the character of dynamics given by completely positive  Davies map. Let us stress, however,  that {\it none} of the physical results reported below depend on the choice of the specifically chosen initial Bell state in Eq. (\ref{bell0}); put differently,  the results remain robust for any of the four Bell states.

\section{Entanglement dynamics}

In this section we investigate entanglement of the two--qubit state $\rho_{AB}(t)$ given in (\ref{evol}).
As a proper measure of entanglement we use the quantifier of {\it negativity} $N(t)$ \cite{hor}, being defined  as
\begin{eqnarray}\label{neg}
N(t) &=&\frac{1}{2}\sum_i \left(|\lambda_i|-\lambda_i\right) \;,
\end{eqnarray}
where the $\lambda_i$ are the eigenvalues of the partially transposed density matrix $\rho_{AB}(t)$, at fixed time $t$,  of a composite system \cite{hor}.

Let us next evaluate this negativity  of the state (\ref{rho}). It explicitly reads:
\begin{eqnarray}\label{negat}
8 N(t) =    2e^{-Gt} +e^{-Ft} - 1 +
\left| 2e^{-Gt} +e^{-Ft} - 1 \right| \;.
\end{eqnarray}
$N(t)$ does not depend on $\omega$,  i.e., it is not affected by the individual single--qubit level spacings in  Eq. (\ref{ham}).
The time evolution of the negativity $N(t)$ is depicted with the upper panel in Fig. \ref{fig2} for $T\to \infty$.
For pure dephasing; i.e. for $F=0$ the negativity exhibits a strictly exponential decay, reading:
\begin{eqnarray}\label{neg_def}
N(t) = \frac{1}{2} {\rm e}^{-G\,t} \;,
\end{eqnarray}
with the characteristic dephasing rate $G=1/\tau_2$.
For a dissipative environment with $ F > 0$,  the  entanglement undergoes a sudden death (ESD) \cite{eberly}
occurring at finite death time $t_c$.  Put differently,  for time $t > t_c$, the entanglement vanished identically.   From Eq. (\ref{negat}) it follows that $t_{c}$ is determined by the relation
\begin{eqnarray}\label{tesd}
 2 e^{-Gt_{c}}+e^{-Ft_{c}} - 1 = 0 \;.
\end{eqnarray}
Upon  increasing the relaxation rate $F$  the death time $t_c$ monotonically decreases from $t_c= \infty$  for $F=0$ to the minimal value
$t_{c}^{min}=  \ln  \left(\sqrt {2} +1 \right)/ G$, which occurs when
$F/G= \tau_2 /\tau_1=2$.
In the case of finite dissipation,  $F > 0$, the negativity $N(t)$--decay proceeds faster than exponentially, see the inset in the upper panel in Fig. 2,  where $N(t)$ {\it vs.} the dimensionless time $Gt$ is depicted on the semi-logarithimic plot with the  vertical axis scaled logarithmically.

\section{Quantum discord}

Next let us  investigate  a quantum correlation measure as encoded with the quantum discord.
We again consider  a composite system consisting of the two qubits  $A$ and $B$.
The full (classical and quantum) correlations in the composite system are encoded with quantum mutual information, defined as
\begin{eqnarray}\label{mut}
{\cal I}(\rho_{AB}) = S(\rho_A)+S(\rho_B)-S(\rho_{AB}),
\end{eqnarray}
where $S(\rho) = -\mbox{Tr}\left(\rho\ln  \rho \right)$ denotes the von-Neumann information entropy, $\rho_{AB}$ is the density operator of the composite bipartite system $AB$. The part  $\rho_{A}$ refers to  the reduced density operator of system part $A$ while, likewise,  $\rho_{B}$ is  the reduced density operator for system  part $B$.

The classical part of the total correlations is defined as a maximum information about one subsystem $A$ that can be obtained by performing a measurement on the other subsystem $B$, as defined  by a complete set of projectors
 $\{\Pi_k^B\}$.
Let us recall that for $p=1/2$, see in Sec. \ref{secII},
an independence of the chosen $A,B$--labelling is granted.
The label $k$  distinguishes different outcomes of this measurement. The quantifier of the classical part of  correlations is defined by the set of following relations
%
%
\begin{eqnarray}\label{klas}
C(\rho_{AB}) = \max_{\{\Pi_k^B\}}S(\rho_{AB}|{\{\Pi^B_k\}}), \\
%
S(\rho_{AB}|{\{\Pi^B_k\}}) =\sum_{k=0}^{1} q_k S(\rho_A^k),  \nonumber\\
\rho_A^k =\frac{1}{q_k}\mbox{Tr}_B\left[\left({\mathbb I}_A\otimes \Pi_k^B  \right) \rho_{AB} \left({\mathbb I}_A\otimes \Pi_k^B  \right)\right], \nonumber  \\
q_k= \mbox{Tr}_{AB}\left[\left({\mathbb I}_A\otimes \Pi_k^B  \right)\rho_{AB} \right].
\end{eqnarray}
%

The difference between the total amount of correlation and the classical part of correlation thus reads
\begin{eqnarray}\label{discord}
D(\rho_{AB}) = {\cal I}(\rho_{AB})-C(\rho_{AB}) \;.
\end{eqnarray}
This relation defines the  {\it quantum discord}. Being so, it provides a measure for manifest quantum correlations \cite{discord}.
Let us remark that, generally, this quantum discord (\ref{discord}) presents neither  a unique nor the most optimal  quantifier for quantum correlation \cite{discord}. However, for the case of bipartite systems one can summarize that  the states can be divided into two groups \cite{rosario};  namely entangled (quantum correlated) and separable states.  In turn, the separable states can either be classically correlated  or quantum correlated (but not entangled).
This classification is non-trivial, as e.g. classically correlated states always lead to completely positive maps while states with quantum correlations may give rise to non-completely positive maps \cite{rosario}.
For our set-up here here, classically correlated states exhibit zero quantum discord $D(t)=0$,  while states with quantum correlations exhibit a non-zero discord $D(t) \geq 0$.

There occurs a natural computational difficulty in evaluating the  quantum discord, stemming from the maximization procedure in Eq. (\ref{klas}).
Fortunately, this task becomes feasible for two qubit systems: it is sufficient to consider projective measurements of the form  \cite{hamieh, discord}
\begin{eqnarray}\label{pomiar}
\Pi_0^B&=&\left(\begin{array}{cc} \cos^2(\theta/2) & \sin(\theta)\exp(i \phi)\\\sin(\theta)\exp(-i \phi) & \sin^2(\theta/2) \end{array} \right),
\nonumber\\
\Pi_1^B&=&{\mathbb I}_B-\Pi_0^B,
\end{eqnarray}
where $\{\theta ,\phi\}$ is a standard parameterization of a single qubit Bloch sphere.
This simplification is helpful in considering the quantum discord for  models which can be effectively described in terms of two qubits
\cite{disdyn,luo,ali}.

The calculation of quantum discord requires, in general, a careful optimization with respect to (being sufficient for us) projective measurements (\ref{pomiar}). Fortunately, here the problem is even more tractable.  Due to symmetry of the system,  finding the optimum  in (\ref{klas}) does not involve the non-trivial $\phi$--dependence,  i.e. the measuring process  (\ref{pomiar}) can be simplified to a single--parameter family of projectors.
The quantum discord therefore is  $\phi$--independent. In addition it exhibits extrema, depending of the value of the dissipation parameter $F$. For $F\le G$,  the maximum in Eq.(\ref{klas}) occurs at $\theta=0$ in Eq.(\ref{pomiar}).  The quantum discord D(t) consequently reads:
%
\begin{eqnarray}\label{dys_0}
4 D(t) &=& (e^{-Ft}+1+2e^{-Gt})\ln(e^{-Ft}+1+2e^{-Gt})\nonumber \\
&+& (e^{-Ft}+1-2e^{-Gt}) \ln(e^{-Ft}+1-2e^{-Gt})\nonumber \\
&-& 2 (1+e^{-Ft}) \ln(1+e^{-Ft}),  \quad
\quad F \le G\;.
\label{d22}
\end{eqnarray}
%
To optimize quantum discord within the required regime $G<F\le 2G$ (see in (\ref{warun})) in turn implies for consistency that $\theta=\pi/2$ in Eq.(\ref{klas}). The final result for $D(t)$ thus reads
%
\begin{eqnarray}\label{dys_pi}
& &4 D(t) = (e^{-Ft}+1+2e^{-Gt}) \ln(e^{-Ft}+1+2e^{-Gt}) \nonumber \\
&+& 2(1-e^{-Ft}) \ln(1-e^{-Ft}) -2(1-e^{-Gt}) \ln(1-e^{-Gt}) \nonumber \\
&+& (e^{-Ft}+1-2e^{-Gt}) \ln(e^{-Ft}+1-2e^{-Gt})  \nonumber \\
&-& 2(1+e^{-Gt}) \ln(1+e^{-Gt}),  \quad \quad
 G<F\le 2G \;.
\label{d23}
\end{eqnarray}
%
%

It follows from Eqs. (\ref{d22}) and (\ref{d23})  that the discord $D(t)$ monotonically approaches $0$  as $t\to \infty$ and stays positive for  finite time, see the lower panel in Fig. 2. Contrary to negativity $N(t)$, which is most robust in the absence of energy relaxation $F=0$, the slowest decay relaxation of quantum discord is  obtained for the case when $F=G$, i.e. when  the two characteristic relaxation times match, i.e.,  $\tau_1 = \tau_2$.  To clarify  this behavior  let us notice that for pure dephasing $F=0$ classical correlations, (\ref{klas})  remain {\it constant} without further decay; -- put differently, Markovian dephasing does not affect the dynamics of classical correlations.

The difference between the quantum discord and the entanglement becomes best visible in the asymptotic long time regime when $t \gg \tau_1$ or $t \gg \tau_2$.
Here, one obtains the limiting behavior
\begin{eqnarray}\label{asym_D}
 D(t) \simeq \frac{e^{-2Gt}}{1+e^{-Ft}} \;,  \quad \mbox{for} \quad F \le G \nonumber \;;  \\
D(t) \simeq \frac{1}{2} \left[ e^{-2Gt} +e^{-2Ft}\right] , \quad \mbox{for} \quad
 G<F\le 2G \;.
\end{eqnarray}
Therefore, for $t\rightarrow \infty$ the decay rate of quantum discord is determined solely by the dephasing time $1/G=\tau_2$ and is independent of the energy relaxation time $\tau_1$. However, the entanglement N(t) vanishes identically in this time-regime if finite energy relaxation with $F>0$ is at work. On the other hand,  for a pure dephasing $F=0$  one finds  the  asymptotic relation
\begin{equation}
D(t) \simeq e^{-2Gt}/2=2 N^2(t)\;.
\label{DNA}
\end{equation}

Although this very  kind of relation between quantum discord $D(t)$ and  entanglement $N(t)$ likely may not be generic  \cite{ali,dyskent}, it nicely illustrates the critical role of an environment for the asymptotic behaviors of both quantifiers. Notably,  the ratio $N(t)/D(t)$ exhibits for $t\rightarrow \infty$ a divergence for a pure
dephasing, while it  vanishes identically in the presence of finite energy relaxation.\\

\section{Conclusions}

With this work we have investigated the time evolution of   entanglement negativity $N(t)$ and  quantum discord $D(t)$ for  a pair of qubits, with one qubit ($B$) weakly coupled to a decohereing  environment. The decoherence dynamics for this sub-system has been approximated by a Markovian, completely positive Davies semi-group dynamics.  We identified two classes of environments that impact differently the quantum correlations as quantified by entanglement and quantum discord: (i) In the  limiting case of the pure decoherence (i.e. strict dephasing) both, the  quantum discord $D(t)$ and the negativity $N(t)$ decay exponentially towards zero in the asymptotic long time regime. Moreover, there exists an appealing  functional relation between these two measures, being detailed with  (\ref{DNA}). (ii) In the case of a dissipative dynamics,  the entanglement undergoes a sudden death at a finite time $t_c$ while the quantum discord $D(t)$ smoothly relaxes towards zero at long times. The slowest  decay of the quantum discord occurs when  the energy relaxation time  $\tau_1$ matches the  dephasing  time $\tau_2$.
Our findings may serve as a potential guideline for the implementation of quantum information, e.g. communication protocols.
We have elucidated which of the two types of correlation measures, namely quantum entanglement $N(t)$ or the one encoded by  quantum discord $D(t)$, can provide an advantageous and/or more suitable quantifier for quantum information  processing occurring in an open quantum systems undergoing an ubiquitous decoherence dynamics.

\section*{Acknowledgement}
This collaborative work has been  supported  by (i) the NCN Grant N202 052940 (JD and  MM), UMO-2011/01/B/ST6/07197 (JD) and (ii) DAAD (J{\L}), as well as by (iii)
the German Excellence Initiative ``Nanosystems Initiative
Munich (NIM)'' (P.H.), (iv) the German collaborative research centre SFB-631 (R.B., P.H.) and by (v) the European Science Foundation
(ESF) under the project "Exploring the Physics of Small Devices" (EPSD).

\end{document}